\documentclass[twocolumn,superscriptaddress,prl]{revtex4-1}
\usepackage{amsmath}
\usepackage{amssymb}
\usepackage{graphicx}
\usepackage{esint}
\usepackage{epstopdf}
\usepackage{amsfonts}
\usepackage{tabularx}
\usepackage{dcolumn}
\usepackage{bm}
\usepackage{ulem}

\makeatletter

\@ifundefined{textcolor}{}
{
 \definecolor{BLACK}{gray}{0}
 \definecolor{WHITE}{gray}{1}
 \definecolor{RED}{rgb}{1,0,0}
 \definecolor{GREEN}{rgb}{0,1,0}
 \definecolor{BLUE}{rgb}{0,0,1}
 \definecolor{CYAN}{cmyk}{1,0,0,0}
 \definecolor{MAGENTA}{cmyk}{0,1,0,0}
 \definecolor{YELLOW}{cmyk}{0,0,1,0}
}

\makeatother

\begin{document}

\title{An experimental proposal to observe non-abelian statistics of Majorana-Shockley
fermions in an optical lattice}

\author{Dong-Ling Deng}
\affiliation{Department of Physics, University of Michigan, Ann Arbor, Michigan 48109, USA}
\affiliation{Center for Quantum Information, IIIS, Tsinghua University, Beijing 100084,
PR China}

\author{Sheng-Tao Wang}
\affiliation{Department of Physics, University of Michigan, Ann Arbor, Michigan 48109, USA}
\affiliation{Center for Quantum Information, IIIS, Tsinghua University, Beijing 100084,
PR China}

\author{Kai Sun}
\affiliation{Department of Physics, University of Michigan, Ann Arbor, Michigan 48109, USA}

\author{Lu-Ming Duan}
\affiliation{Department of Physics, University of Michigan, Ann Arbor, Michigan 48109, USA}
\affiliation{Center for Quantum Information, IIIS, Tsinghua University, Beijing 100084,
PR China}

\begin{abstract}
We propose an experimental scheme to observe non-abelian statistics
with cold atoms in a two dimensional optical lattice. We show that the Majorana-Schockley
modes associated with line defects obey non-abelian statistics and can
be created, braided, and fused, all through adiabatic shift of
the local chemical potentials. The detection of the topological qubit is transformed
to local measurement of the atom number on a single lattice site. We demonstrate the robustness of
the braiding operation by incorporating noise and experiential imperfections in numerical simulations,
and show that the requirement fits well with the current experimental technology.
\end{abstract}

\maketitle

Besides the conventional bosons and fermions, in synthetic
two-dimensional (2D) materials there exist more exotic quasi-particles
with non-abelian statistics, meaning that the state of the system
will be transformed by non-commutable unitary operators when we adiabatically
braid the particles one around the other \cite{stern2010non}. Search
for such non-abelian particles is one of the hottest topics in physics
\cite{nayak2008non,stern2010non,wilczek2009majorana,alicea2012new,beenakker2013search}.
Observation of the non-abelian statistics is
of both fundamental interest and practical importance, in particular
for topological quantum computation \cite{kitaev2003fault,nayak2008non}
Despite the recent great progress \cite{mourik2012signatures,das2012zero,rokhinson2012fractional,veldhorst2012josephson,alicea2012new,beenakker2013search},
it remains technically elusive to braid the quasi-particles in materials
to verify their conjectured non-abelian statistics \cite{stern2010non}.

Laser controlled cold atoms provide a powerful experimental platform
to realize exotic states of matter \cite{dalibard2011colloquium,bloch2012quantum,lewenstein2012ultracold}.
Several proposals have been made to observe non-abelian statistics
based on control of vortex states in a $p+ip$ superfluid \cite{sato2009non,tewari2007quantum,zhu2011probing}.
A vortex in a $p+ip$ superfluid of odd vorticity traps a zero energy
mode corresponding to a Majorana fermion, which is its own antiparticle
and described by a real fermion operator \cite{wilczek2009majorana}.
The Majorana fermions in different vortices are found to obey non-abelian
statistics \cite{read2000paired,ivanov2001non,stern2004geometric}.
An intriguing proposal has been made to braid the vortex Majorana
fermions in cold atomic gas by a focused laser beam \cite{tewari2007quantum,zhu2011probing}.
An experimental implementation of this proposal, however, is still
challenging for several reasons: first, besides the Majorana mode
a vortex traps a number of other states which have a small gap to
the zero-energy mode \cite{tewari2007quantum}. This small gap sets
a tough requirement for the relevant energy and time scales. Second,
moving of the vortex by a focused laser beam may change its trapped
modes, and a quantitative understanding of this process is still lacking.
Finally, a natural way to realize the $p+ip$ superfluid is based
on the $p$-wave Feshbach resonance \cite{gurarie2005quantum}, but
the latter is difficult to stabilize in free space \cite{gaebler2007p,han2009stabilization}.
Very recently, another nice idea has been suggested to braid Majorana
modes associated with dislocations in an optical lattice \cite{Buhler2014Majorana}. Insertion
of dislocations requires change of structure of the optical lattice,
which is experimentally challenging and yet to be demonstrated.

In this paper, we propose an experimental scheme to observe non-abelian
statistics with cold atoms in an optical lattice in a vortex-free
configuration. A $p$-wave superfluid based on the Feshbach resonance
could be stabilized in an optical lattice due to the quantum Zeno
effect \cite{syassen2008strong,han2009stabilization}. The recent
remarkable experimental advance has allowed single-site addressing
in a 2D optical lattice \cite{sherson2010single,weitenberg2011single,bakr2009quantum,bakr2010probing,Wurtz2009Experimental}.
By this ability, we can create a line defect in a 2D lattice simply
by shifting the chemical potential along the line. Different from
the dislocations, this line defect requires no change of structure
of the optical lattice and is ready to be implemented in current experiments
\cite{sherson2010single,weitenberg2011single,bakr2009quantum,bakr2010probing,Wurtz2009Experimental}.
It was found recently that a pair of zero-energy modes emerge at the
edges of this line defect \cite{wimmer2010majorana} by the Shockley mechanism \cite{shockley1939surface}.
The exchange statistics of these modes, however, remains unresolved
\cite{wimmer2010majorana}. Motivated by recent works on braiding
of nanowires \cite{alicea2011non}, here we show through exact numerical
simulation that the Majorana-Shockley modes associated with these
line defects in a 2D superfluid obey non-abelian statistics and their
braiding can be achieved by tuning of only the local chemical potential.
This tuning is significantly simpler compared with the braiding of
nanowires \cite{alicea2011non} or dislocations \cite{Buhler2014Majorana}, which requires
site-by-site tuning of the pairing interaction and the tunneling rates
\cite{alicea2011non,kraus2013braiding}. We demonstrate robustness
of the braiding operation under influence of practical noise and propose
a scheme to measure the topological qubits using local measurement
of the atom number. The proposed scheme fits well with the state-of-the-art
of the experimental technology in a 2D optical lattice \cite{sherson2010single,weitenberg2011single,bakr2009quantum,bakr2010probing,Wurtz2009Experimental}.

\noindent We consider cold atoms in a 2D optical lattice, which are
prepared into the $p+ip$ superfluid phase. This superfluid phase
can be achieved, for instance, through the $p$-wave Feshbach resonance
\cite{gurarie2005quantum}, which leads to the $p+ip$ superfluid
phase under a wide range of experimental parameters \cite{gurarie2005quantum}.
The instability associated with the $p$-wave Feshbach resonance in
free space \cite{gaebler2007p} could be overcome in an optical lattice
through the quantum Zeno effect \cite{han2009stabilization}. Alternatively,
an effective $p+ip$ superfluid phase for cold atoms can also be achieved
by a combination of the $s$-wave Feshbach resonance and the light
induced spin-orbital coupling \cite{fu2008superconducting,sau2010generic,zhu2011probing}.

In the momentum $\mathbf{k}$ space, the Bogoliubov-de Gennes (BdG)
Hamiltonian describing the $p+ip$ superfluid phase on a square optical
lattice has the form $H=\sum_{\mathbf{k}}\psi_{\mathbf{k}}^{\dagger}\mathcal{H}(\mathbf{k})\psi_{\mathbf{k}}$,
with $\psi_{\mathbf{k}}^{\dagger}=(c_{\mathbf{k}}^{\dagger},c_{-\mathbf{k}})$
and
\begin{equation}
\mathcal{H}(\mathbf{k})=d_{x}(\mathbf{k})\sigma^{x}+d_{y}(\mathbf{k})\sigma^{y}+d_{z}(\mathbf{k})\sigma^{z},\label{eq:Hamiltonian}
\end{equation}
where $d_{x}(\mathbf{k})=\Delta\sin k_{x}a,$ $d_{y}(\mathbf{k})=\Delta\sin k_{y}a,$
$d_{z}(\mathbf{k})=\mu-J(\cos k_{x}a+\cos k_{y}a),$ $\sigma^{x,y,z}$
denote the Pauli matrices, $a$ is the lattice constant, $\mu$ is
the chemical potential, $J$ is the neighboring hopping rate, and
$\Delta$ is the pairing interaction strength. The topological property
of this Hamiltonian is characterized by the first Chern number $C_{1}=-\frac{1}{2\pi}\int_{\text{BZ}}dk_{x}dk_{y}F_{xy}(\mathbf{k})$
with the Berry curvature $F_{xy}(\mathbf{k})=\partial_{k_{x}}A_{y}(\mathbf{k})-\partial_{k_{y}}A_{x}(\mathbf{k})$
and the Berry connection $A_{\nu}(\mathbf{k})=\langle u_{-}(\mathbf{k})|i\partial_{k_{\nu}}|u_{-}(\mathbf{k})\rangle$
($\nu=x,y$), where $|u_{-}(\mathbf{k})\rangle$ denotes the lower
band Bloch eigenstate of $\mathcal{H}(\mathbf{k})$ and the integration
in $C_{1}$ is over the first Brillouin zone (BZ). The phase of the
Hamiltonian $H$ is topologically nontrivial with $C_{1}=$sign$(\mu)$
in the parameter regime $0<|\mu|<2J$ (taking $\Delta$ as the energy
unit) and topologically trivial with $C_{1}=0$ when $|\mu|>2J$.
A topological phase transition occurs at $|\mu|=2J$.

\begin{figure}
\includegraphics[width=0.49\textwidth]{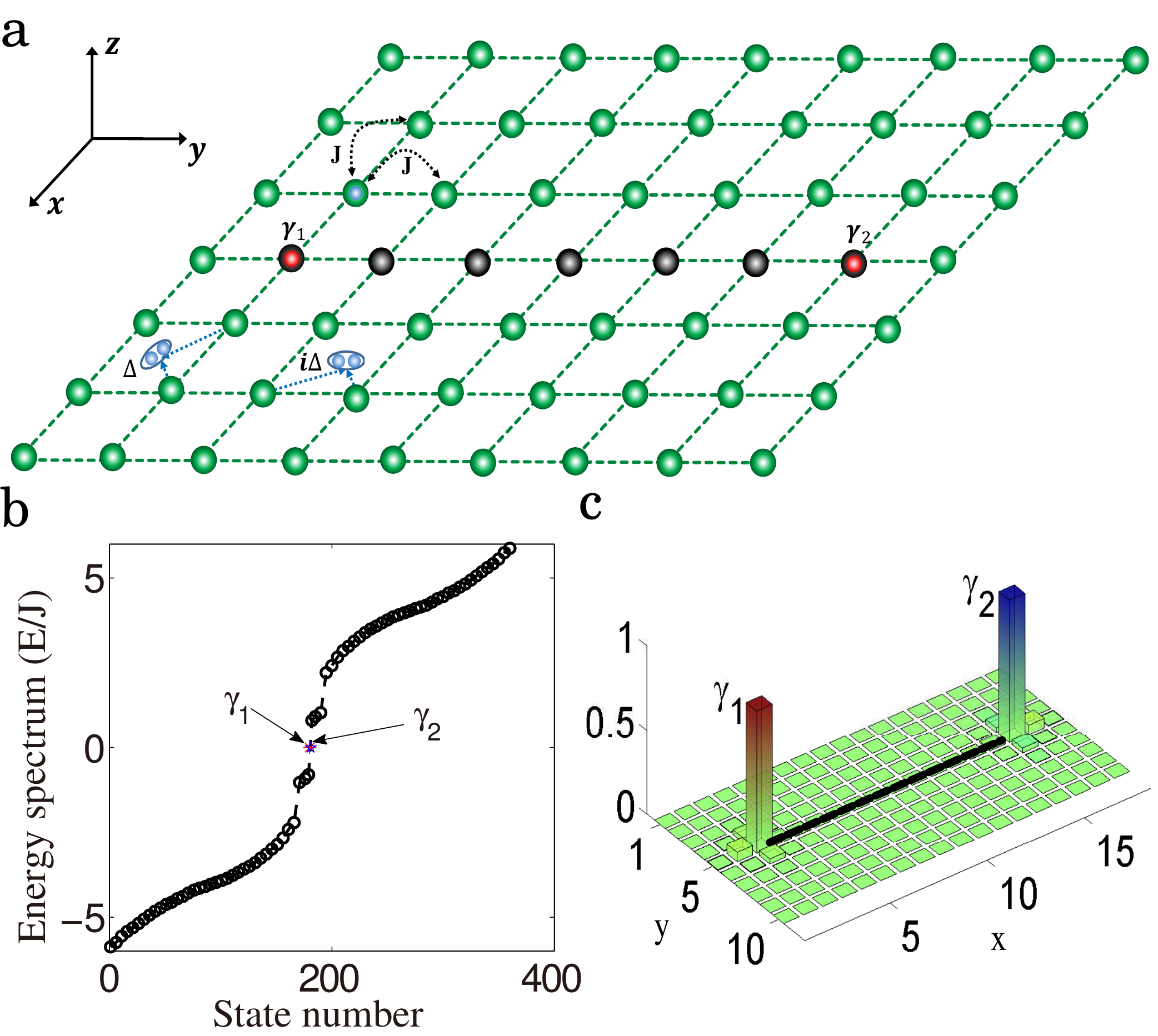}
\caption{\textbf{Creation and manipulation of the MSFs in an optical lattice.}
(\textbf{a}) Cold fermionic atoms are loaded into a 2D optical lattice.
$J$ and $\Delta$ denotes the nearest neighbor hopping rate and pairing
strength. A line defect with different local chemical potential binds
two zero-energy MSFs $\gamma_{1}$ and $\gamma_{2}$ (red circles)
at its edges. (\textbf{b}) Energy spectrum of the Hamiltonian $H$
on a square lattice of size $18a\times10a$ with open boundaries.
The length of the line defect is $14a$. The zero-energy MSFs have
tiny energy splitting due to the small size of the line defect, which
is numerically found to be $<10^{-10}J$ for our parameters. (\textbf{c})
The amplitude of the mode function for $\gamma_{1}$ and $\gamma_{2}$.
The black line indicates the line defect with chemical potential $\mu_{d}$.
The parameters are chosen as $\Delta=J$, $\mu_{0}=10J$, and $\mu_{d}=0.1J.$}
\label{OpLEn}
\end{figure}

\noindent With single-site addressing, the potential shift of each
lattice site can be individually adjusted in experiments \cite{sherson2010single,weitenberg2011single,bakr2009quantum,bakr2010probing,Wurtz2009Experimental}.
We create a line defect in a 2D optical lattice by tuning the chemical
potential $\mu_{d}$ along a chain of atoms to make it different from
that of the background lattice (denoted by $\mu_{0}$) so that they
reside in topologically distinct phases (illustrated in Fig. 1a).
For a certain range of $\mu_{d}$ that depends on $\mu_{0}$, a pair
of zero energy Majorana-Shockley fermion (MSF) modes appear at the
two edges of the line defect \cite{wimmer2010majorana}. We choose
$\mu_{0}$ in the topologically trivial phase with $\mu_{0}>2J$ so
that there are no other zero-energy modes on the boundary of the finite
2D lattice.

\begin{figure}
\includegraphics[width=0.49\textwidth]{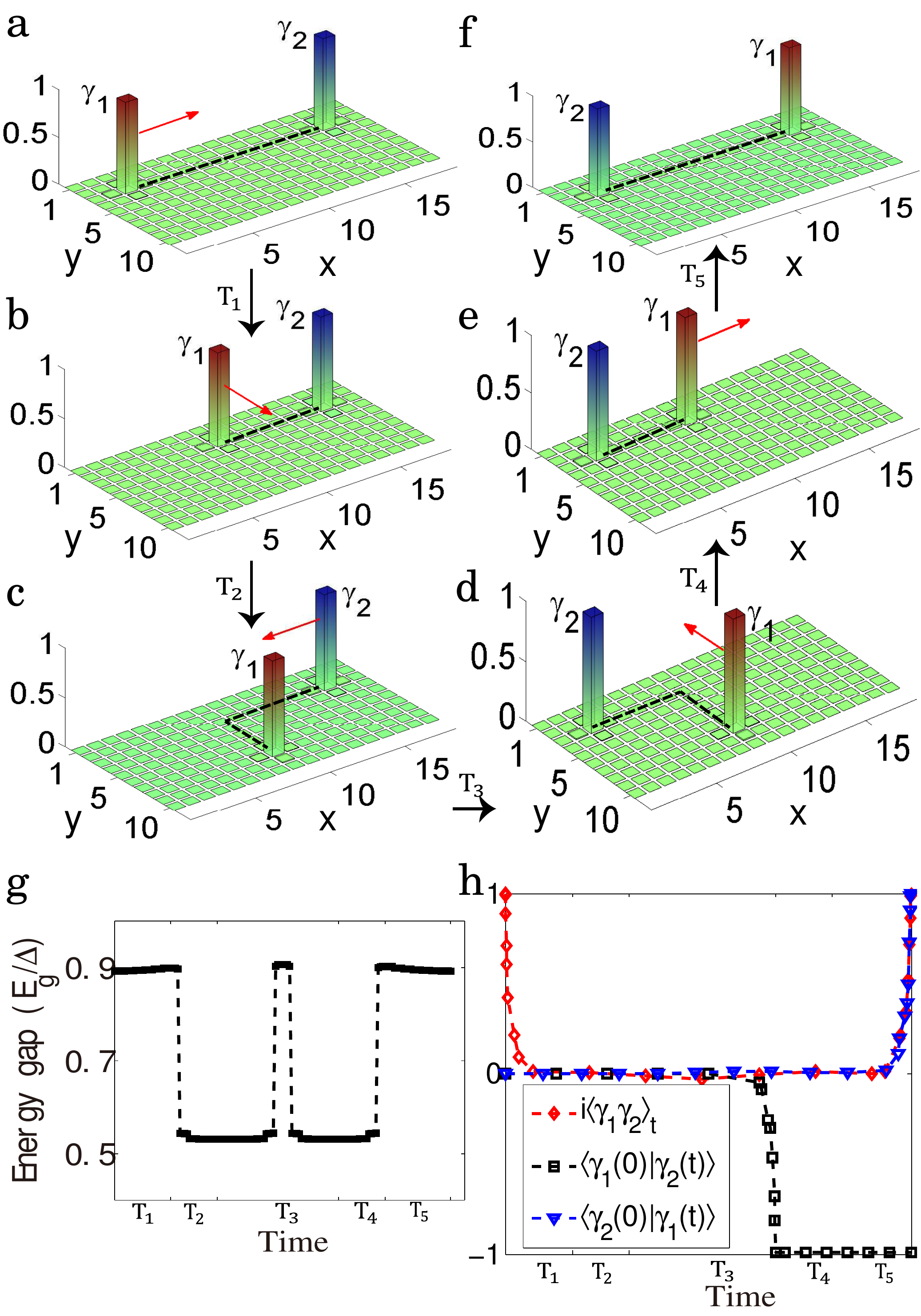}
\caption{\textbf{Braiding of two MSFs bound to the same line defect.} (\textbf{a})
The black line indicates the line defect with chemical potential $\mu_{d}$.
Sequentially tuning the local chemical potentials at one end from
$\mu_{d}$ to $\mu_{0}$ shortens the line defect and transports $\gamma_{1}$
along the $x$ direction. The red arrow shows the moving direction
of the MSF. Similar operations along a T-junction path realize adiabatic
exchange of $\gamma_{1}$ and $\gamma_{2}$, with steps illustrated
in (\textbf{a})-(\textbf{f}). (\textbf{g}) The evolution of the energy
gap $E_{g}$ throughout the braiding process. The system is always
gapped with the minimum gap $E_{g}>0.5J$. (\textbf{h}) Time evolution
of the MSF modes $\gamma_{1},\gamma_{2}$ and their correlations.
All the parameters are the same as in Fig.\ \protect \ref{OpLEn}. }
\label{fig:Braiding-two-Majorana}
\end{figure}

Under a typical size of the 2D optical lattice with a line defect,
we solve exactly the eigenmodes of the Hamiltonian (1) under the open
boundary condition, and the eigen-spectrum is shown in Fig.\ \ref{OpLEn}b
(see Methods). Clearly, there are a pair of zero-energy MSF modes
that are separated from other defect modes and bulk states by a minimum
gap about $J$. The MSFs are described by anti-commuting real fermion
operators $\gamma_{j}$ with $\gamma_{j}=\gamma_{j}^{\dag}$ and $\gamma_{j}\gamma_{k}+\gamma_{k}\gamma_{j}=2\delta_{jk}$.
A pair of MSF modes $\gamma_{1}$ and $\gamma_{2}$ together represent
a conventional fermion mode $c_{m}=\left(\gamma_{1}+i\gamma_{2}\right)/2$,
with the eigenstates of $c_{m}^{\dag}c_{m}=i\gamma_{1}\gamma_{2}+1$
encoding a topological qubit. The eigen-functions of the MSF modes
$\gamma_{1}$ and $\gamma_{2}$ are shown in Fig. 1c, which are well
localized at the edges of the line defect.

To examine the exchange statistic of the MSF modes, we adiabatically
deform the line defect with steps shown in Fig. 2(a-f). Each step
is achieved through site-by-site tuning of the chemical potential
from $\mu_{d}$ to $\mu_{0}$ (to shorten the line defect) or from
$\mu_{0}$ to $\mu_{d}$ (to extend the line defect). We simulate
the time evolution of the MSF modes in the Heisenberg picture under
adiabatic evolution of the Hamiltonian. The Hamiltonian remains gapped
at any time as shown in Fig. 2g, which protects the MSF modes from
mixing of the other modes. The evolution of the MSF modes $\gamma_{1}$
and $\gamma_{2}$ and their correlation are shown in Fig. 2h. After
the whole evolution with time $T$, apparently we have $\gamma_{1}\left(T\right)=\gamma_{2}(0)$
and $\gamma_{2}\left(T\right)=-\gamma_{1}(0)$. The correlation $\left\langle \gamma_{1}\left(T\right)\gamma_{2}\left(T\right)\right\rangle =-\left\langle \gamma_{2}\left(0\right)\gamma_{1}\left(0\right)\right\rangle =\left\langle \gamma_{1}\left(0\right)\gamma_{2}\left(0\right)\right\rangle $.

\begin{figure}
\includegraphics[width=0.49\textwidth]{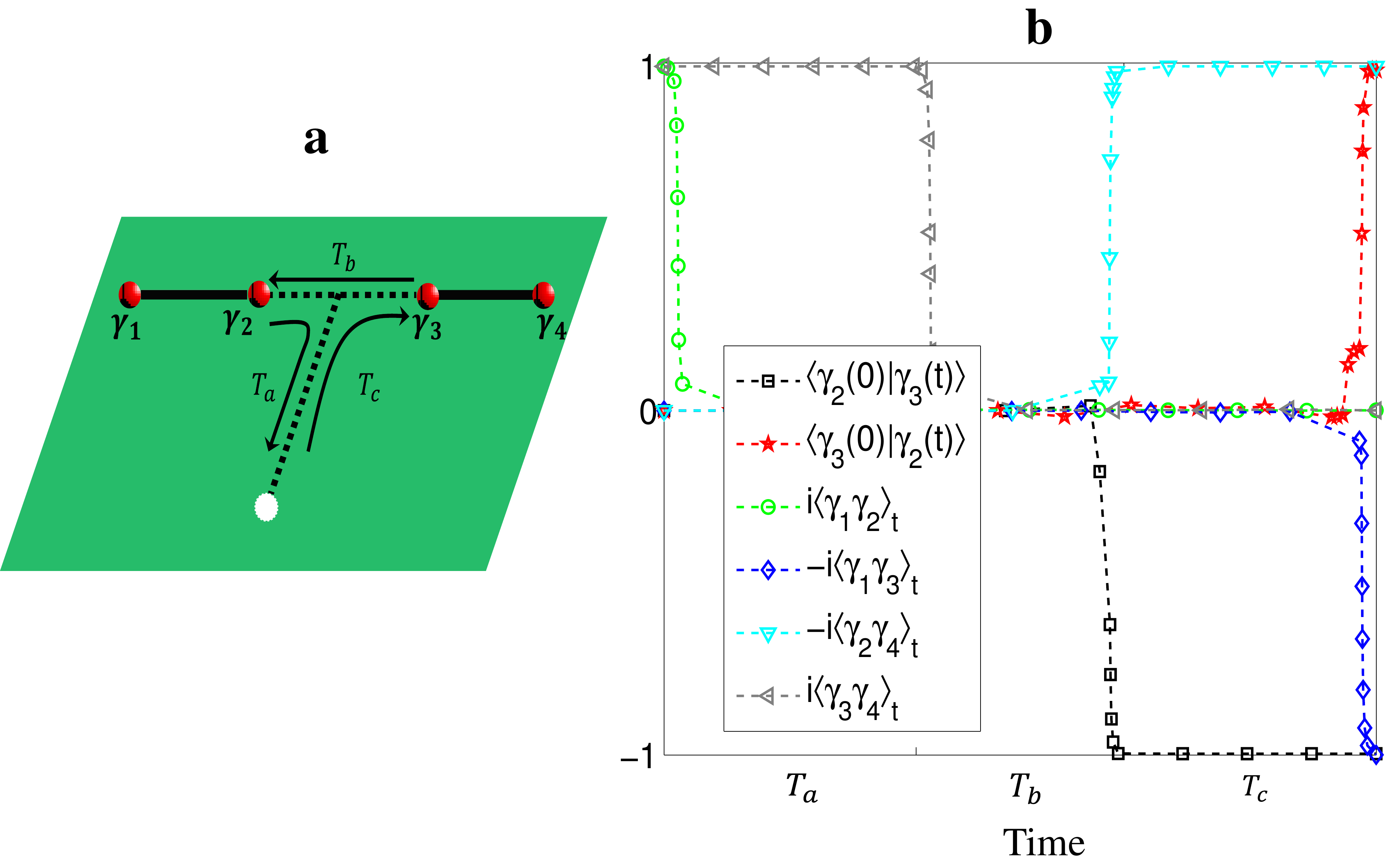}
\caption{\textbf{Braiding of two MSFs bound to different line defects.} (\textbf{a})
Illustration of braiding two MSFs from different line defects along
the T-junction path. (\textbf{b}) Time evolution of the MSF modes
$\gamma_{1},\gamma_{2},\gamma_{3},\gamma_{4}$ and their correlations.
The MSFs $\gamma_{2}$ and $\gamma_{3}$ are braided. The parameters
are taken as: the lattice size $12a\times28a$, two horizontal line
defects each of length $8a$ and distance $9a$, $\Delta=0.91J$,
$\mu_{0}=10\Delta$ and $\mu_{d}=0.1\Delta$. }
\label{fig:Time-evolution}
\end{figure}

\noindent This transformation of the MSF modes occurs in a similar
way when we adiabatically braid the edges associated with different
line defects. In Fig. 3a, we illustrate the adiabatic braiding of
two edge modes $\gamma_{2}$ and $\gamma_{3}$ of different line defects
along a T-junction path. This braiding involves joining and cutting
of two line defects and we need to choose parameters appropriately
to avoid appearance of accidental near-zero-energy modes. In general,
the four zero-energy MSF modes are still well protected by a significant
energy gap. Their evolution and the associated correlations are shown
in Fig. 3b. The results indicate that $\gamma_{2}\left(T\right)=\gamma_{3}(0)$
and $\gamma_{3}\left(T\right)=-\gamma_{2}(0)$ for the two braided
modes. The other modes remain unchanged with $\gamma_{1}\left(T\right)=\gamma_{1}(0)$
and $\gamma_{4}\left(T\right)=\gamma_{4}(0)$.

The above transformation rule generalizes straightforwardly to the
case of $2N$ MSF modes. The rule is exactly the same as the case
of Majorana fermions bound to vortices. For $2N$ modes $\gamma_{j}$
($j=1,2,\cdots,2N$), when we braid $\gamma_{j}$ and $\gamma_{j+1}$,
the transformation is described by a unitary operator $U_{j}=e^{\pi\gamma_{j+1}\gamma_{j}/4}$
which transforms $\gamma_{j}\rightarrow\gamma_{j+1}$,$\gamma_{j+1}\rightarrow-\gamma_{j}$.
As $U_{j}$ and $U_{j+1}$ do not commute, the exchange statistics
of the MSF modes are non-abelian and belongs to the so-called Ising
anyon class according to classification of non-abelian anyons \cite{di1997conformal}.

The unitary operation $U_{j}$ from topological braiding of the MSF
modes are robust to noise and experimental imperfections. To test
that, we consider several sources of noise typical for atomic experiments:
First, with imperfect single-site addressing, when we tune the chemical
potential of one site, we may change the potentials of the neighboring
sites as well, modeled by a spreading ratio of $1-\alpha$. Second,
there is a global weak harmonic trap for cold atom experiments, with
an additional trapping potential $V_{\text{trap}}=\frac{V_{\text{T}}}{2(L_{x}^{2}+L_{y}^{2})}\sum_{\mathbf{r}}d_{\mathbf{r}}^{2}c_{\mathbf{r}}^{\dagger}c_{\mathbf{r}}$,
where $L_{x}$ ($L_{y}$) is the lattice dimension along the $x$
($y$) direction, and $d_{\mathbf{r}}$ is the distance from the trap
center. Typically, $V_{\text{T}}$ ranges from $0.1J$ to $J$. Finally,
there is unavoidable small disorder potential in experiments which
adds random fluctuation to the chemical potential with magnitude denote
by $\lambda_{R}$. We recalculate the evolution of the MSF modes and
their correlation, incorporating contribution of all these sources
of noise. The results are shown in Fig. 4, which are almost indistinguishable
from the corresponding results shown in Fig. 2h under the ideal case.
This demonstrates robustness of the braiding operations of the MSFs.

To verify the non-abelian braiding operations, we need to detect the
topological qubit encoded by two nonlocal MSF modes $\gamma_{1}$
and $\gamma_{2}$. For the 1D nanowire, the parity of the total particle
number is a conserved property, which is different for the two eigenstates
of $i\gamma_{1}\gamma_{2}$ and thus can be used to detect the topological
qubit \cite{alicea2011non,kraus2013braiding}. For our case, the line
defect interacts with the background lattice with tunneling and pairing
terms which in general do not conserve the parity of the total atom
number along the line, therefore the parity detection does not work.
We propose a different method to detect the topological qubit. The
line defect is adiabatically shortened until it finally reduces to
a single lattice site $\mathbf{r}_{0}$ (illustrated in Fig. 5a) and
we examine evolution of the MSF modes $\gamma_{1}$ and $\gamma_{2}$
during this process. As shown in Fig. 5b, with a high fidelity ( about
$99\%$), the mode $\gamma_{1}$ ($\gamma_{2}$) is mapped to $\gamma_{\mathbf{r}_{0},A}=c_{\mathbf{r}_{0}}^{\dagger}+c_{\mathbf{r}_{0}}$
($\gamma_{\mathbf{r}_{0},B}=i(c_{\mathbf{r}_{0}}^{\dagger}-c_{\mathbf{r}_{0}})$),
respectively. By a measurement of the local atom number $c_{\mathbf{r}_{0}}^{\dagger}c_{\mathbf{r}_{0}}$
after the adiabatic merging, we thus measure the topological operator
$i\gamma_{1}\gamma_{2}$ with a high fidelity (about $98\%$). This
local measurement is actually more robust compared with the nonlocal
parity detection. Note that the detection fidelity of the topological
qubit in principle can be improved to an arbitrary accuracy by using
the quantum non-demolition (QND) technique: to measure the topological
qubit $i\gamma_{1}\gamma_{2}$, we create an ancillary topological
qubit (with MSF modes $\gamma_{3}$ and $\gamma_{4}$), perform an
effective Controlled-NOT gate between the topological qubits $i\gamma_{1}\gamma_{2}$
and $i\gamma_{3}\gamma_{4}$ through the noise-resilient braiding
operations \cite{Deng2013Fault-tolerant}., and then measure the ancilla
$i\gamma_{3}\gamma_{4}$ by the above method. As the qubit $i\gamma_{1}\gamma_{2}$
is not destroyed by the measurement, it can be repeatedly measured
through this QND\ technique and the detection error is exponentially
suppressed with increase of the detection rounds.

\begin{figure}
\includegraphics[width=0.38\textwidth]{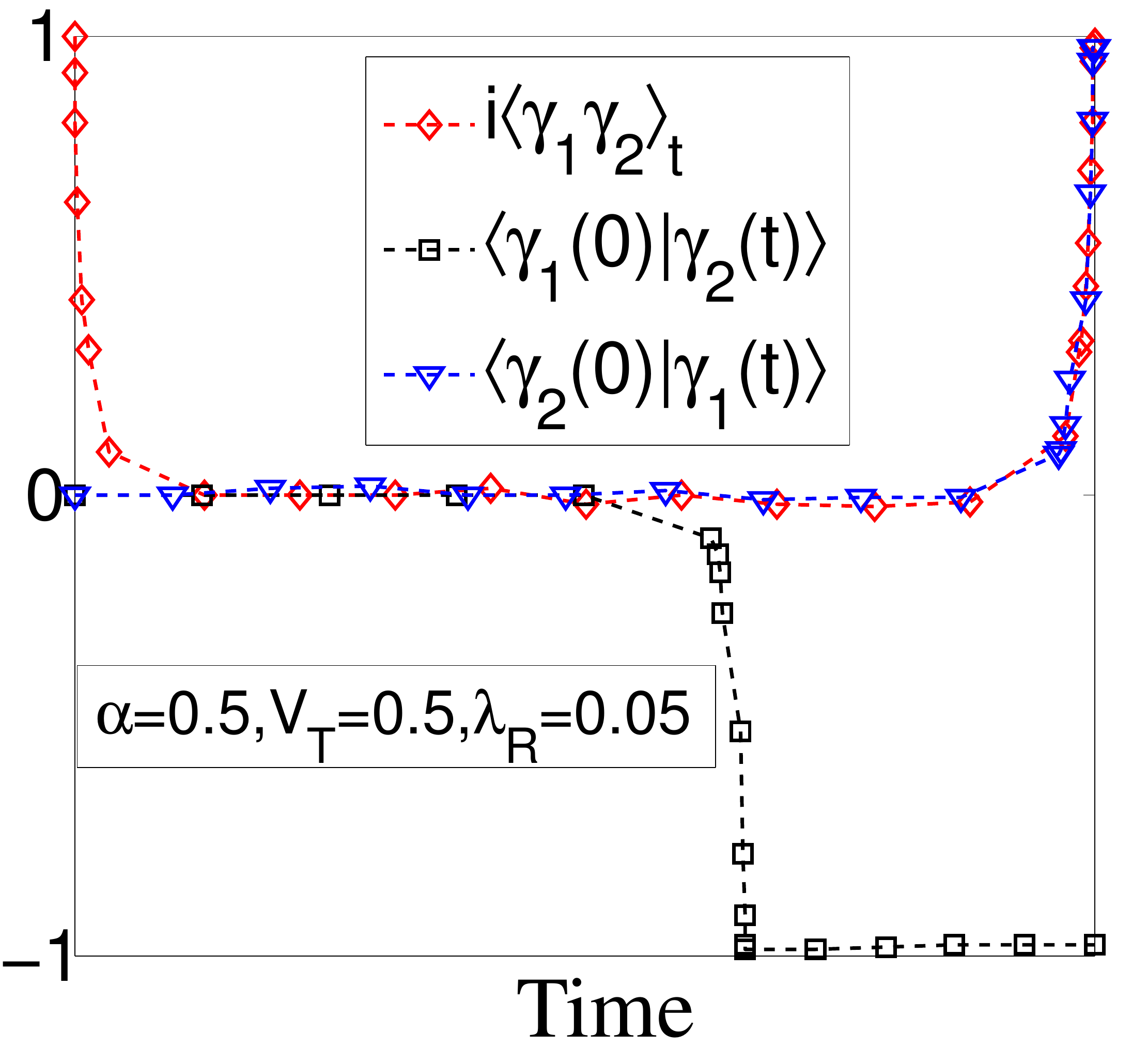}
\caption{\textbf{Robustness to experimental noise and imperfections.} The lattice
size is $20a\times12a$ and other parameters are the same as in Fig. \protect\ref{OpLEn}.
$\alpha,V_{T},\lambda_{R}$ denote the parameters characterizing respectively
the laser beam crosstalk, the strength of the global harmonic trap,
and the magnitude of random fluctuation of the chemical potential
(see the main text). }
\label{fig:Robustness-to-realistic}
\end{figure}

\noindent
\begin{figure}
\includegraphics[width=0.49\textwidth]{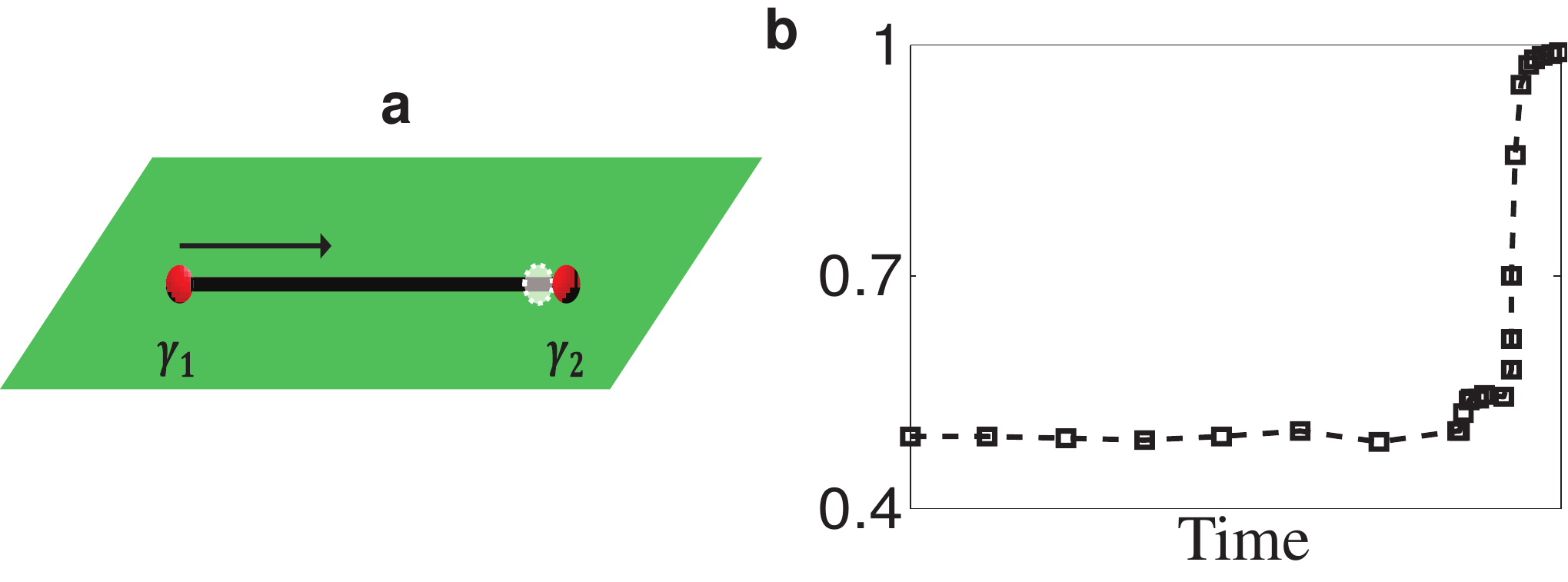}
\caption{\textbf{Detection of the topological qubit.} (\textbf{a}) Two MSFs
$\gamma_{1}$ and $\gamma_{2}$ are fused through adiabatic shortening
of the line defect to a single lattice site $\mathbf{r}_{0}$. (\textbf{b})
Transformation of the MSF modes $\gamma_{1}$ and $\gamma_{2}$ under
adiabatic merging. For simplicity, we plot evolution of the magnitude
of the mode overlap between $c_{\mathbf{r}_{0}}$ and $[\gamma_{1}(t)+i\gamma_{2}(t)]/2$.
At the end of merging, $\gamma_{1}$ and $\gamma_{2}$ are mapped
dominantly to the local modes $c_{\mathbf{r}_{0}}^{\dagger}+c_{\mathbf{r}_{0}}$
and $i(c_{\mathbf{r}_{0}}^{\dagger}-c_{\mathbf{r}_{0}})$, respectively,
which enables detection of the initial nonlocal topological qubit
by a simple measurement of the atom number $c_{\mathbf{r}_{0}}^{\dagger}c_{\mathbf{r}_{0}}$
on a single lattice site after the adiabatic merging. All the parameters
are the same as in Fig. \ref{OpLEn}. }
\label{fig:Measures}
\end{figure}

In summary, we have proposed a complete scheme to observe non-abelian
statistics of the MSFs associated with line defects in a 2D optical
lattice. The MSFs are created, braided, and fused all through adiabatic
tuning of the chemical potential for certain lattice sites. The detection
of the topological qubit is transformed to local measurement of the
atom number on a single lattice site. The required technology well
fits with the current status of the optical lattice experiments \cite{sherson2010single,weitenberg2011single,bakr2009quantum,bakr2010probing,Wurtz2009Experimental}.
Through numerical simulation, we have demonstrated robustness of the
non-abelian braiding operations under typical experimental configuration
with imperfections. The scheme provides a viable approach for observation
of the exotic non-abelian braiding statistics, which is a goal of
intense interest and a critical step for realization of robust topological
quantum information processing \cite{kitaev2003fault,nayak2008non,Deng2013Fault-tolerant}.

\noindent \textbf{Appendix}

\textbf{Time evolution.} We first Fourier transform the Hamiltonian
(1) into real space, with the modes in real space denoted by $c_{\mathbf{r}}$.
A line defect has chemical potential $\mu_{d}$ instead of $\mu_{0}$.
We define the Majorana operators at each lattice site with $\gamma_{\mathbf{r},A}=(c_{\mathbf{r}}^{\dagger}+c_{\mathbf{r}})$
and $\gamma_{\mathbf{r},B}=i(c_{\mathbf{r}}^{\dagger}-c_{\mathbf{r}})$.
In terms of these Majorana operators, the Hamiltonian has the following
form:
\begin{eqnarray}
H=\frac{i}{2}\sum_{\mathbf{p},\mathbf{q}}\mathcal{H}_{\mathbf{p}\mathbf{q}}\gamma_{\mathbf{p}}\gamma_{\mathbf{q}},
\end{eqnarray}
where $p=(\mathbf{r},\beta)$ and $q=(\mathbf{r}^{\prime},\beta^{\prime})$
($\beta,\beta^{\prime}=A,B$) are combined indices and $\mathcal{H}$
is a $2N\times2N$ real skew-symmetric matrix with $N$ being the
number of lattice sites.

By locally and adiabatically tuning $\mu$ along a T-junction path,
MSFs can be braided. During this process, the Majorana operators evolve
according to the following equation in the Heisenberg picture \cite{kraus2007quantum}:
\begin{equation}
\gamma_{\mathbf{p}}\rightarrow\gamma_{\mathbf{p}}(t)=U\gamma_{\mathbf{p}}(0)U^{\dagger}=\sum_{\mathbf{q}}\mathcal{O}_{\mathbf{q}\mathbf{p}}\gamma_{\mathbf{q}}\left(0\right),\label{eq:time-evolution}
\end{equation}
where $U=\mathcal{T}\exp[i\int_{0}^{t}H(\tau)d\tau]$ and $\mathcal{O}=\mathcal{T}\exp[-i\int_{0}^{t}\mathcal{H}(\tau)d\tau]$
is an element of the special orthogonal group $\mathcal{O}\in\text{SO}(2N)$;
$\mathcal{T}$ is the time-ordering operator.

In our numerical simulation, we first diagonalize $\mathcal{H}$ at
time $t=0$ to obtain the zero-energy eigen modes $\gamma_{i}(0)=\sum_{\mathbf{p}}\eta_{i\mathbf{p}}\gamma_{\mathbf{p}}(0)$,
where the coefficients $\eta_{i\mathbf{p}}$ represent the mode function
and are localized at the ends of the line defects. During the braiding
process, the zero-energy eigen modes evolve as $\gamma_{i}(t)=U\gamma_{i}(0)U^{\dagger}=\sum_{\mathbf{p}}\eta_{i\mathbf{p}}\gamma_{\mathbf{p}}(t)$,
where $\gamma_{\mathbf{p}}(t)$ are calculated via Eq. (\ref{eq:time-evolution}).
Using this method, we obtain the time evolution of the zero-energy
MSF modes with the results plotted in the main text.

\noindent \textbf{Majorana correlation functions}. To calculate the
Majorana correlations, we use the method introduced in Ref.\ \cite{kraus2009pairing}.
Let us define the density operator $\rho=N\exp(-\beta H)$ ($N$ is
the normalization constant and $\beta$ is the inverse temperature)
and the antisymmetric covariance matrix $\Gamma$ with elements $\Gamma_{\mathbf{p}\mathbf{q}}=\frac{i}{2}\text{Tr}[\rho(\gamma_{\mathbf{p}}\gamma_{\mathbf{q}}-\gamma_{\mathbf{q}}\gamma_{\mathbf{p}})]$.
The Hamiltonian $H$ can be brought into block off-diagonal form $OHO^{\text{T}}=\oplus_{j=1}^{N}\left(\begin{array}{cc}
0 & -\epsilon_{j}\\
\epsilon_{j} & 0
\end{array}\right)$ by a special orthogonal matrix $O\in\text{SO}(2N)$, where $\epsilon_{j}$
characterizes the energy eigen-spectrum of the Hamiltonian. This matrix
$O$ also reduces $\Gamma$ to a block off-diagonal form $O\Gamma O^{\text{T}}=\oplus_{j=1}^{N}\left(\begin{array}{cc}
0 & \eta_{j}\\
-\eta_{j} & 0
\end{array}\right)$ with $\eta_{j}=\tanh(\beta\epsilon_{j}/2)$. The covariance matrix
$\Gamma_{\text{G}}$ corresponding to the ground state of $H$ is
obtained by letting the inverse temperature $\beta\rightarrow\infty$,
i.e., $\eta_{j}\rightarrow\text{sgn}(\epsilon_{j})$. After we obtain
$\Gamma_{\text{G}}$, the Majorana correlations can be computed by
Wick's theorem via the equation:
\begin{eqnarray}
i\langle\gamma_{\mathbf{p}}\gamma_{\mathbf{q}}\rangle=\text{Pf}(\Gamma_{\text{G}}^{^{\prime}}),
\end{eqnarray}
where $\Gamma_{\text{G}}^{^{\prime}}=\left(\begin{array}{cc}
(\Gamma_{\text{G}})_{\mathbf{p\mathbf{p}}} & (\Gamma_{\text{G}})_{\mathbf{p\mathbf{q}}}\\
(\Gamma_{\text{G}})_{\mathbf{q\mathbf{p}}} & (\Gamma_{\text{G}})_{\mathbf{q\mathbf{q}}}
\end{array}\right)$ is a $2\times2$ submatrix of $\Gamma_{\text{G}}$ and $\text{Pf}(\Gamma_{\text{G}}^{^{\prime}})$
is the Pfaffian of $\Gamma_{\text{G}}^{^{\prime}}$ with $\text{Pf}(\Gamma_{\text{G}}^{^{\prime}})^{2}=\det(\Gamma_{\text{G}}^{^{\prime}})$.
Once we have obtained $i\langle\gamma_{\mathbf{p}}\gamma_{\mathbf{q}}\rangle$
at time $t=0$, the time evolution of the MSF mode correlations $i\langle\gamma_{i}\gamma_{j}\rangle_{t}$
can be computed directly using $\gamma_{i}(t)=\sum_{\mathbf{p}}\eta_{i\mathbf{p}}\gamma_{\mathbf{p}}(t)=\sum_{\mathbf{p,q}}\eta_{i\mathbf{p}}\mathcal{O}_{\mathbf{q}\mathbf{p}}\gamma_{\mathbf{q}}\left(0\right)$.

\noindent \textbf{Acknowledgement} We thank J. Alicea, C. V. Kraus, and Y.-H Chan for discussions.
D.L.D., S.T.W., and L.M.D. are supported by the NBRPC (973 Program) 2011CBA00300 (2011CBA00302),
the IARPA MUSIQC program, the ARO and the AFOSR MURI program. K.S. is supported in part by NSF PHY-1402971.

\noindent \textbf{Author contributions }All the authors contribute
substantially to this work.

\noindent \textbf{Competing financial interests} The authors declare
no competing financial interests.

\end{document}